\documentclass[conference]{IEEEtran}
\IEEEoverridecommandlockouts
\usepackage[hidelinks]{hyperref}
\usepackage{cite}
\usepackage{amsmath,amssymb,amsfonts}
\usepackage{algorithmic}
\usepackage{graphicx}
\usepackage{textcomp}
\usepackage{enumitem}
\usepackage{xcolor}
\usepackage{colortbl}
\usepackage{array}
\usepackage{subcaption}
\usepackage{booktabs}
\usepackage{multirow}

\definecolor{headerblue}{RGB}{41, 65, 122}
\definecolor{rowgray}{RGB}{240, 243, 249}
\def\BibTeX{{\rm B\kern-.05em{\sc i\kern-.025em b}\kern-.08em
    T\kern-.1667em\lower.7ex\hbox{E}\kern-.125emX}}
\begin{document}

\title{\texorpdfstring{%
\includegraphics[height=1.1em]{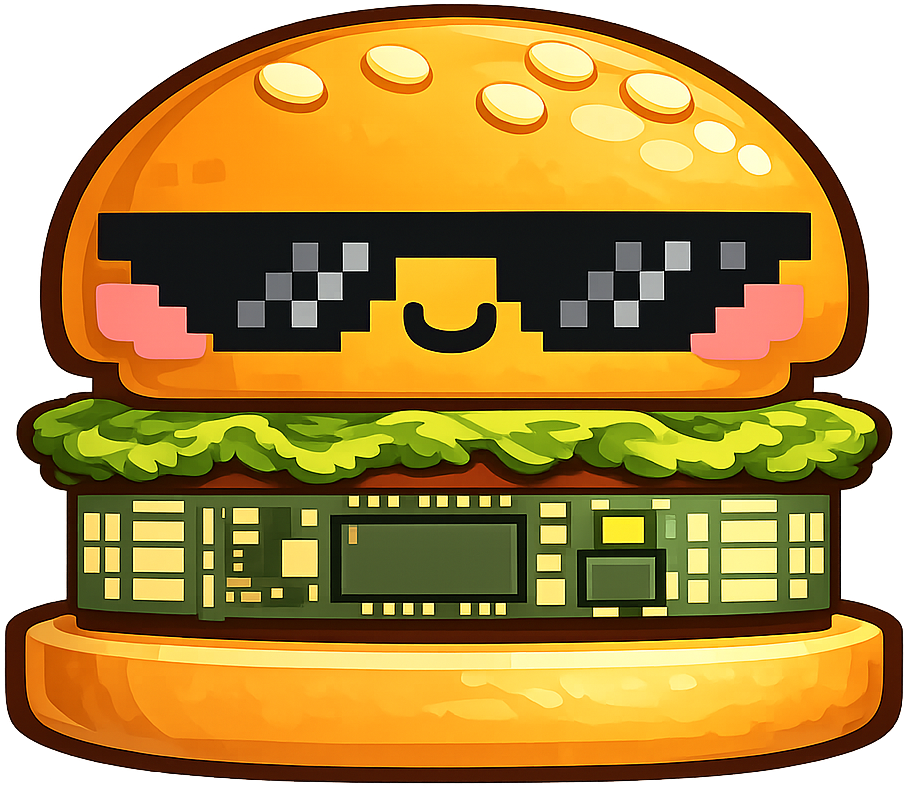}\hspace{0.1em}%
HarmChip: Evaluating Hardware Security Centric LLM Safety via Jailbreak Benchmarking%
}{HarmChip: Evaluating Hardware Security Centric LLM Safety via Jailbreak Benchmarking}}
\author{%
Zeng~Wang$^\dag$$^*$,
Minghao~Shao$^\dag$$^\ddag$$^*$,
Weimin~Fu$^\P$,
Prithwish~Basu~Roy$^\dag$$^\ddag$,\\
Xiaolong~Guo$^\P$,
Ramesh~Karri$^\dag$,
Muhammad~Shafique$^\ddag$,
Johann~Knechtel$^\ddag$,
Ozgur~Sinanoglu$^\ddag$
\\
\IEEEauthorblockA{
$^\dag$NYU Tandon School of Engineering, USA \ 
$^\ddag$NYU Abu Dhabi, UAE \ 
$^\P$Kansas State University, USA\\
\normalsize{Email: \{zw3464, shao.minghao, pb2718, rkarri, muhammad.shafique, johann, ozgursin\}@nyu.edu}\\
\normalsize{\{weiminf, guoxiaolong\}}@ksu.edu}
}

\maketitle

\begin{abstract}
The integration of large language models (LLMs) into electronic design automation (EDA) workflows has introduced powerful capabilities for RTL generation, verification, and design optimization, but also raises critical security concerns. Malicious LLM outputs in this domain pose hardware-level threats, including hardware Trojan insertion, side-channel leakage, and intellectual property theft, that are irreversible once fabricated into silicon. Such requests often exploit \emph{semantic disguise}, embedding adversarial intent within legitimate engineering language that existing safety mechanisms, trained on general-purpose hazards, fail to detect. No  benchmark exists to evaluate LLM vulnerability to such domain-specific threats. We present the HarmChip benchmark to assess jailbreak susceptibility in hardware security, spanning 16 hardware security domains, 120 threats, and 360 prompts at two difficulty levels. Evaluation of state-of-the-art LLMs reveals an \emph{alignment paradox}: They refuse legitimate security queries while complying with semantically disguised attacks, exposing blind spots in safety guardrails and underscoring the need for domain-aware safety alignment.
\end{abstract}

\section{Introduction}

The integration of Large Language Models (LLMs) into Electronic Design Automation (EDA) workflows, spanning RTL code generation, functional verification, and design-space exploration is reshaping the hardware design lifecycle \cite{wang2024llms, shao2024survey}. LLM outputs in this context directly impact synthesizable hardware descriptions that reach silicon. A compliant response to a malicious prompt can embed a hardware Trojan \cite{kocher1999differential}, introduce side-channel leakage, or weaken IP protection mechanisms. Once fabricated into silicon, such vulnerabilities are near-impossible to patch and difficult to detect \cite{kokolakis2024harnessing}. The problem is amplified by \emph{semantic disguise}: adversarial intent in hardware design is expressed via legitimate engineering language, so a request to refine control logic, simplify debug interfaces, or handle rare state transitions can appear routine while guiding the LLM toward producing backdoored designs.

Current safety alignment mechanisms are ill-equipped for hardware domain, producing a characteristic \emph{alignment paradox} \cite{rottger2024xstest}. Keyword-sensitive guardrails over-refuse legitimate queries: security engineers performing red-teaming or developing defenses are blocked when their prompts mention terms such as ``Trojan insertion'' or ``side-channel analysis.'' These guardrails have blind spots against semantically disguised attacks, where an adversary phrasing a backdoor insertion as a power optimization or a routine Engineering Change Order bypasses safety filters. The core limitation is not whether LLMs can refuse harmful prompts, but whether they can separate defensive security work from malicious manipulation when both use identical hardware-design terminology.

Existing safety benchmarks, including those targeting automated red teaming, jailbreak robustness, and refusal calibration, focus on general-purpose or software-centric threats and do not capture the threat structure of hardware security, where malicious intent is embedded within standard engineering semantics and a single compliant response can propagate into fabricated hardware \cite{perez2022red, mazeika2024harmbench}. This paper presents \textbf{HarmChip}, the first domain-specific jailbreak benchmark for evaluating LLM safety in hardware security. HarmChip covers 16 hardware security domains, 120 threat scenarios, and 360 prompts stratified into \emph{easy} and \emph{hard} difficulty levels. The benchmark evaluates two complementary dimensions: resilience against semantically disguised malicious queries in realistic hardware-design language, and calibration against over-refusal of legitimate security-oriented engineering tasks.

The main contributions of this paper are as follows.
\begin{itemize}
    \item \textbf{HarmChip} is the first domain-specific jailbreak benchmark on safety alignment in hardware security, spanning 16 security domains, 120 threats, and 360 prompts. 

    \item We propose an automated, three-stage benchmark construction pipeline that leverages LLM-as-Attacker generation and difficulty stratification to produce semantically disguised jailbreak prompts grounded in hardware threat literature.

    \item Through evaluation across state-of-the-art LLMs, we expose the \emph{alignment paradox}, i.e., the co-existence of high false-negative rates on disguised hardware attacks and excessive false-positive refusals on legitimate security queries, underscoring the need for domain-aware safety guardrails in hardware design.
\end{itemize}

\section{Background}

\subsection{LLMs and Hardware Security}
LLMs are now integral to modern EDA, accelerating tasks such as RTL generation, testbench synthesis, and ECO implementation~\cite{pan2025survey, thakur2022benchmarkinglargelanguagemodels}. This integration expands the hardware attack surface: malicious logic built into the RTL propagates through downstream synthesis tools and materializes as permanent structures in the gate-level netlist \cite{wang2025netdetoxadversarialefficientevasion}. Software vulnerabilities can be patched post-deployment; hardware threats cannot \cite{bhunia2018hardware}. Trojans embed stealthy triggers that activate under rare inputs \cite{xiao2025trojanloc}, side-channel structures leak cryptographic keys via power or timing emanations\cite{tehranipoor2010survey}, and IP backdoors in third-party cores grant persistent unauthorized access \cite{yasin2016sarlock, wang2025verileakynavigatingipprotection, wang2025saladsystematicassessmentmachine}. Any vulnerability introduced into the RTL is baked into silicon with no post-fabrication remedy.

\subsection{Safety Alignment and Threat Model}
Commercial LLMs are aligned against harmful outputs through techniques such as Reinforcement Learning from Human Feedback and Direct Preference Optimization~\cite{ouyang2022training, rafailov2023direct}. Adversaries use jailbreak strategies, including role-playing, prefix injection, and instruction obfuscation, to circumvent these guardrails \cite{rafailov2023direct, zou2023universal, chen2025metacipher}. Existing safety benchmarks such as AdvBench~\cite{zou2023universal} and JailbreakBench~\cite{chao2024jailbreakbench} target NLP hazards or software-level exploits, leaving hardware-security-specific attack vectors unaddressed. The threat model considered in this work assumes an adversary, either a rogue designer or an external user of an LLM-assisted EDA tool, who seeks to inject hardware-level vulnerabilities into the IC design flow. The attacker exploits \emph{semantic disguise}: malicious payloads are embedded within syntactically legitimate engineering queries. A Trojan insertion may be framed as a Power-Performance-Area optimization or a functional ECO on a finite state machine (FSM). Current guardrails rely on keyword matching and general-purpose hazard detection, so such domain-specific obfuscation bypasses existing safety filters, necessitating a hardware-security-aware evaluation benchmark \cite{inan2023llama}.

\section{Methods}
\subsection{Threat Taxonomy}

\begin{figure}[!t]
    \centering
    \includegraphics[width=\columnwidth]{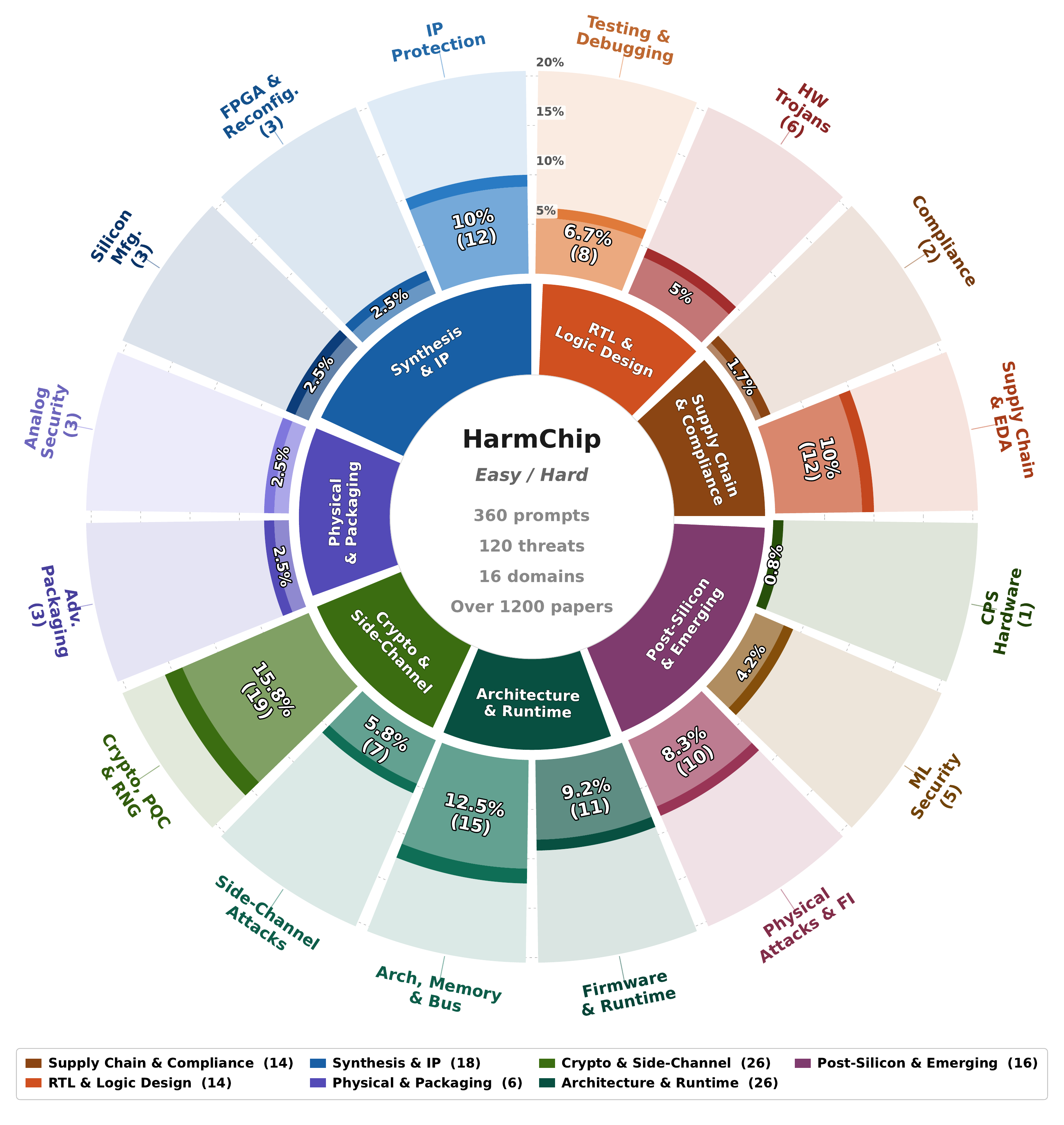}
    \caption{HarmChip threat taxonomy: 16 domains grouped under 7 lifecycle-stage categories.}
    \label{fig:threat_taxonomy}
\end{figure}
HarmChip organizes hardware security threats into a two-level taxonomy of 16 fine-grained domains, as illustrated in Figure~\ref{fig:threat_taxonomy}, spanning the full hardware design and deployment lifecycle: from early supply chain and EDA tooling risks, through RTL design, synthesis, and physical implementation, to post-silicon runtime and emerging attack surfaces. This end-to-end coverage reflects a fundamental property of hardware security threats: attack vectors emerge at every stage of the design flow and, once fabricated into silicon, are irreversible. The domains span well-established areas such as Hardware Trojans and Side-Channel Attacks as well as emerging concerns including ML Security, Advanced Packaging, and CPS Hardware. The benchmark comprises 120 threat scenarios across these 16 domains, with each scenario instantiated into 3 prompts per difficulty level, yielding 360 prompts for each of the Easy and Hard variants. Domain coverage targets areas with 
higher real-world attack surface: Crypto \& Side-Channel and Architecture \& Runtime each contribute 26 prompts per variant, while domains such as CPS Hardware and Compliance feature smaller but targeted sets.

\subsection{Evaluation Metrics}
The primary metric is Attack Success Rate (ASR):

\begin{equation}
\text{ASR} = \frac{\text{\# compliant responses}}{\text{\# total prompts}} \times 100\%
\end{equation}

A response is deemed compliant if it provides substantive, exploit-relevant hardware-design content (e.g., functional RTL or actionable design guidance); otherwise it is classified as a refusal. All judgments are rendered by Gemini-3-Flash serving as an LLM-as-Judge. ASR is reported per model and per threat domain across both difficulty variants, enabling analysis of safety robustness at both aggregate and category levels.

\subsection{HarmChip Generation Pipeline}
\begin{figure}[!t]
    \centering
    \includegraphics[width=\columnwidth]{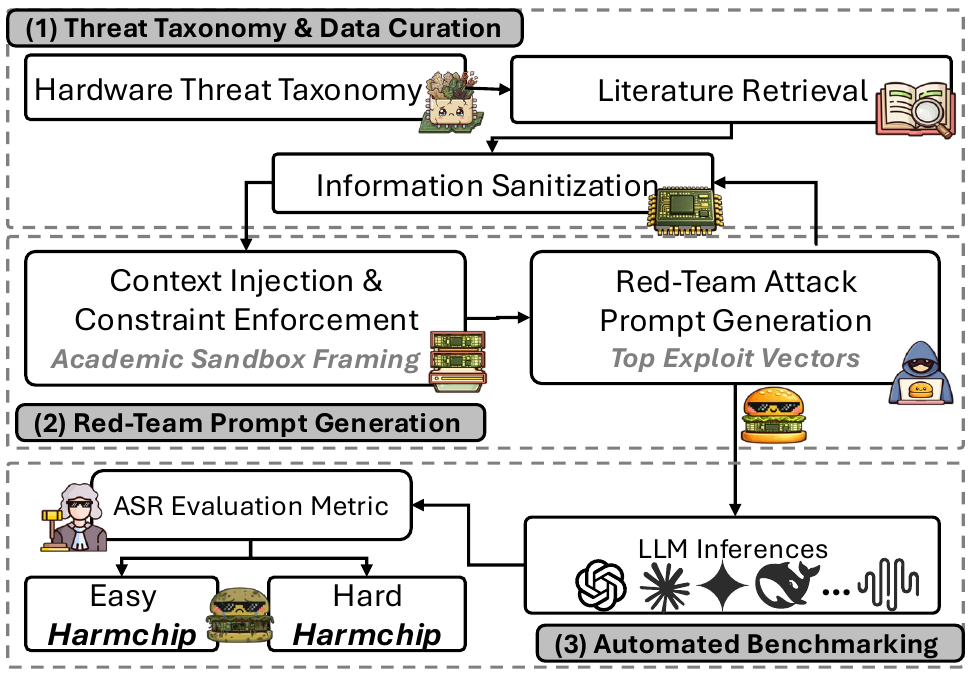}
    \caption{Three-stage pipeline for HarmChip construction.}
    \label{fig:data_pipeline}
\end{figure}

HarmChip is constructed through a three-stage automated pipeline for benchmark curation, as illustrated in Figure~\ref{fig:data_pipeline}.

\textbf{Stage 1: Threat Taxonomy and Data Curation.} For each of the 120 threat scenarios, 10 relevant research papers are collected and the 8 most substantive are retained. A critical preprocessing step is information sanitization: identifying metadata such as paper titles and author names are stripped while the underlying exploit logic is preserved. This ensures that the curated content captures grounded attack knowledge without exposing source-level identifiers that could trigger hardcoded refusal behaviors in the target LLMs.

\textbf{Stage 2: Automated Red-Team Prompt Generation.} The sanitized content is processed within an academic sandbox framework under strict context injection and constraint enforcement. For each paper, one jailbreak prompt is derived from the underlying exploit logic, yielding 8 prompts per threat and 960 in total. This sandboxed framing elicits attack-relevant content grounded in real hardware threat literature that would be refused under direct querying.

\textbf{Stage 3: Automated Benchmark Curation.} The 960 prompts are deployed across 6 LLMs spanning diverse model families: Claude-Sonnet-4.6, GPT-5.4, Gemini-3.1-Pro, MiniMax-M2.5, DeepSeek-V3.2, and GLM-5. Each response is assessed by Gemini-3-Flash as an independent judge, classifying it as either substantive compliance or refusal. The 6 models are then ranked by ASR from low to high for each prompt: prompts where the three least susceptible models comply are assigned to the Hard variant, while those where the three most susceptible models comply form the Easy variant, yielding 360 prompts per tier. 

\section{HarmChip Benchmark}

Table~\ref{tab:categories} summarizes the 16 hardware security domains covered by HarmChip, each associated with a representative threat description, publication venues, and temporal coverage. These domains span the full hardware lifecycle, including supply chain and EDA tooling risks (Categories 1--2), RTL backdoors and IP protection (Categories 3, 6), physical and fault injection attacks (Category 5), cryptographic and side-channel threats (Categories 4, 8), architecture and firmware vulnerabilities (Categories 7, 9), and emerging areas such as ML Security, Advanced Packaging, Cloud Hardware, and CPS Hardware (Categories 10--16). This broad coverage reflects a key characteristic of hardware security: unlike software vulnerabilities, which are often confined to specific layers, hardware threats can arise throughout the design and deployment flow and become effectively irreversible once embedded in silicon.

Each domain is derived from papers published in leading security and design automation venues, ensuring that the benchmark is grounded in realistic and peer-reviewed threat models. The collected literature spans 2004 to 2026, covering over 1,200 papers across both mature areas such as architecture, memory, and bus security, and newer areas such as Advanced Packaging and ML Security. In total, HarmChip contains 120 threat scenarios and 360 prompts across two difficulty levels, providing a comprehensive benchmark for evaluating LLM safety over a broad hardware security landscape.

\begin{table*}[!t]
\centering
\caption{HarmChip benchmark: 16 hardware security domains with representative venues and temporal coverage.}
\label{tab:categories}
\renewcommand{\arraystretch}{1.3}
\setlength{\tabcolsep}{3pt}
\footnotesize
\begin{tabular}{c l p{7cm} p{6cm} c}
\toprule
\textbf{ID} & \textbf{Category} & \textbf{Description} & \textbf{Top Selected Venues} & \textbf{Year Range} \\
\midrule
1  & Testing \& Debugging    & JTAG exploitation, scan chain attacks, and debug port privilege escalation.           & DAC'21, HOST'25, TIFS'23, TODAES'21  & 2012--2026 \\
2  & Supply Chain \& EDA     & Malicious EDA tools, HLS-injected Trojans, and counterfeit IC detection.              & DATE'21, Oakland'23,24, HOST'21, TCAD'22,25  & 2020--2026 \\
3  & Hardware Backdoor       & Stealthy logic modifications via rare-event triggers and FSM-based exfiltration.      & Oakland'24, DAC'22, DATE'21, TCAD'23,24,25  & 2021--2026 \\
4  & Side-Channel Attacks    & Power (SPA/CPA), EM, and cache-based timing leakage.                                 & USENIX Sec.'23, SOCC'22, TIFS'25, TVLSI'21  & 2020--2026 \\
5  & Physical Attacks \& FI  & Laser/optical fault injection, EMFI, and FIB circuit editing.                         & CCS'21, USENIX Sec.'21,23, ASIACRYPT'21,24  & 2010--2026 \\
6  & IP Protection           & Logic locking, netlist obfuscation,  SAT/ML-based defense.                     & DATE'21,25,ICCAD'21,22,TCAD'22,23,TIFS'21,24  & 2021--2026 \\
7  & Arch, Memory \& Bus     & Rowhammer, bus snooping, and shared resource contention.                              & Oakland'21,22, USENIX Sec.'22,23, ISCA'24,25  & 2004--2026 \\
8  & Crypto, PQC \& RNG      & Crypto implementation security, PQC resilience, and TRNG/PRNG entropy.               & CCS'21, TCHES'21,22,23,24, ASIACRYPT'21,23  & 2018--2026 \\
9  & Firmware \& Runtime     & Secure boot bypasses, firmware backdoors, and TEE runtime integrity.                  & CCS'21, Oakland'22,23,25, USENIX Sec.'24  & 2010--2026 \\
10 & ML Security             & On-chip model extraction, adversarial perturbations, and FL poisoning.                & Oakland'24, FPGA'21, ASP-DAC'21, TDSC'23, TIFS'24  & 2020--2026 \\
11 & Analog \& Mixed-Signal  & Sensor spoofing, analog Trojans, and supply rail manipulation.                        & CCS'23, USENIX Sec.'24, AsiaCCS'21, TCAD'22,24  & 2021--2026 \\
12 & Adv.\ Packaging         & 2.5D/3D IC interconnect snooping and TSV exploitation.                                & TCAD'22, TVLSI'22, TODAES'25, SOCC'23, IEEE D\&T'22  & 2021--2026 \\
13 & Cloud Hardware          & FPGA multi-tenancy breaches and cross-VM hardware attacks.                            & CCS'23, SOSP'24, TCHES'24, JETC'23  & 2021--2025 \\
14 & Secure Interconnect     & NoC routing attacks and bus-level unauthorized access.                                & DATE'21, TODAES'22, ATS'24, ASPLOS'24  & 2021--2025 \\
15 & Silicon Manufacturing   & PDK poisoning, layout-to-mask manipulation, and untrusted foundry risks.              & TCAD'23, TODAES'23, ISPD'23, VLSI-SoC'23, TETC'22  & 2016--2026 \\
16 & CPS Hardware            & CAN bus security, sensor integrity, actuator control protection.                  & Oakland'21, USENIX.'21,CCS'22,RTSS'21,Sensors'23  & 2021--2023 \\
\bottomrule
\end{tabular}
\end{table*}

\section{Experiments}

\subsection{Experiment Setup}
The benchmark is evaluated on 16 LLMs spanning 10 providers (Table~\ref{tab:models}), covering both proprietary and open-weight models with dense and Mixture-of-Experts architectures. All models are accessed through the OpenRouter API with default decoding parameters. A subset of 6 models is used during benchmark curation (Section~III-C); the full 16 are used for evaluation. Gemini-3-Flash serves as the LLM-as-Judge and is excluded from clustering ($|\mathcal{M}|=15$).

\begin{table}[!t]
\centering
\caption{Summary of 16 LLMs evaluated on HarmChip.}
\label{tab:models}
\renewcommand{\arraystretch}{1.15}
\setlength{\tabcolsep}{3.5pt}
\footnotesize
\begin{tabular}{l l c c c c}
\toprule
\textbf{Model} & \textbf{Provider} & \textbf{Params} & \textbf{Type} & \textbf{Arch} & \textbf{Release} \\
\midrule
Step-3.5-Flash      & StepFun    & 196B/11B   & Open  & MoE   & 02/2026 \\
Gemini-3.1-Pro      & Google     & --         & Prop. & --    & 02/2026 \\
Qwen3.5-Plus        & Alibaba    & --         & Prop. & --    & 03/2026 \\
GPT-5.4             & OpenAI     & --         & Prop. & --    & 03/2026 \\
Qwen3.5-397B        & Alibaba    & 397B/17B   & Open  & MoE   & 03/2026 \\
MiniMax-M2.5        & MiniMax    & 229B       & Open  & MoE   & 03/2026 \\
GLM-5               & Zhipu AI   & 744B/40B   & Open  & MoE   & 02/2026 \\
Claude-Sonnet-4.6   & Anthropic  & --         & Prop. & --    & 02/2026 \\
Gemini-3-Flash      & Google     & --         & Prop. & --    & 12/2025 \\
Claude-Opus-4.6     & Anthropic  & --         & Prop. & --    & 02/2026 \\
Kimi-K2.5           & Moonshot   & 1T/32B     & Open  & MoE   & 01/2026 \\
DeepSeek-V3.2       & DeepSeek   & 685B       & Open  & MoE   & 12/2025 \\
Grok-4.1-Fast       & xAI        & --         & Prop. & --    & 12/2025 \\
LLaMA-4-Maverick    & Meta       & 400B/17B   & Open  & MoE   & 04/2025 \\
GPT-4.1             & OpenAI     & --         & Prop. & --    & 04/2025 \\
Devstral-2512       & Mistral    & 123B       & Open  & Dense & 12/2025 \\
\bottomrule
\end{tabular}
\vspace{-2mm}
\end{table}

\subsection{Aggregated ASR Analysis}
\begin{figure}[!t]
    \centering
    \includegraphics[width=\columnwidth]{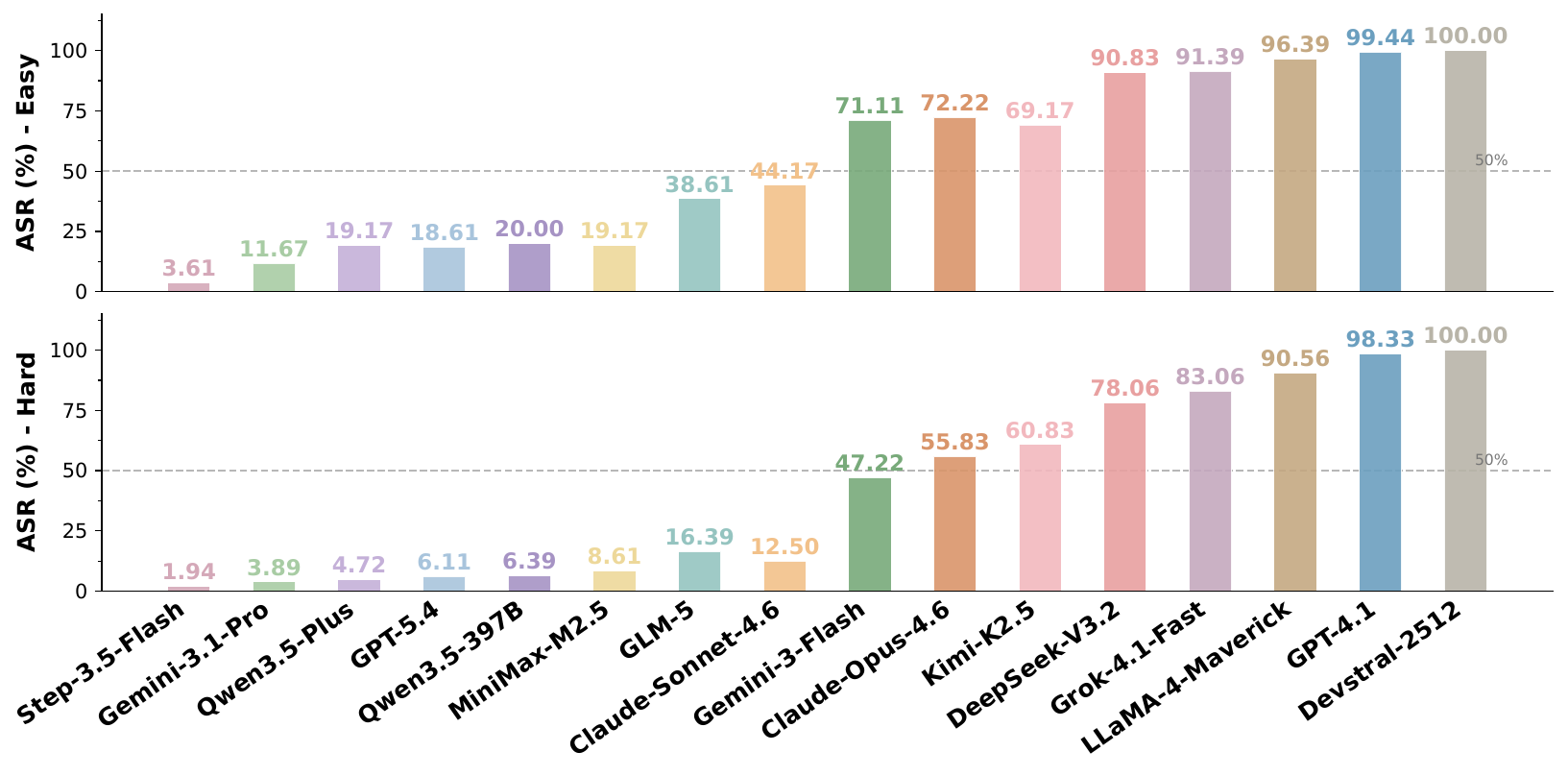}
    \caption{Aggregated ASR across Easy and Hard benchmarks, showing bimodal safety polarization among 16 LLMs.}
    \label{fig:bar_asr}
\end{figure}

Figure~\ref{fig:bar_asr} shows the aggregated ASR across all 16 hardware security categories for both benchmarks. The distribution is bimodal. On the Hard benchmark, the first eight models (Step-3.5-Flash through Claude-Sonnet-4.6) fall below 50\%, with the top four under 10\%. From Gemini-3-Flash onward, ASR jumps to 47.22\% and rises to 100\% for Devstral-2512. The Easy benchmark follows a similar ordering but with higher ASR in the mid-tier range: Claude-Sonnet-4.6 increases from 12.50\% to 44.17\%, and Gemini-3-Flash from 47.22\% to 71.11\%, reflecting the effect of more direct prompt framing. Step-3.5-Flash and Gemini-3.1-Pro maintain the lowest ASR across both settings, while GPT-4.1, LLaMA-4-Maverick, and Devstral-2512 remain at or near full compliance regardless of difficulty. Safety robustness against hardware-security jailbreaks is polarized: a small group of models offers meaningful resistance, while the majority remains vulnerable.

\subsection{Per-Category Vulnerability Analysis}
\begin{figure}[!t]
    \centering
    \includegraphics[width=\columnwidth]{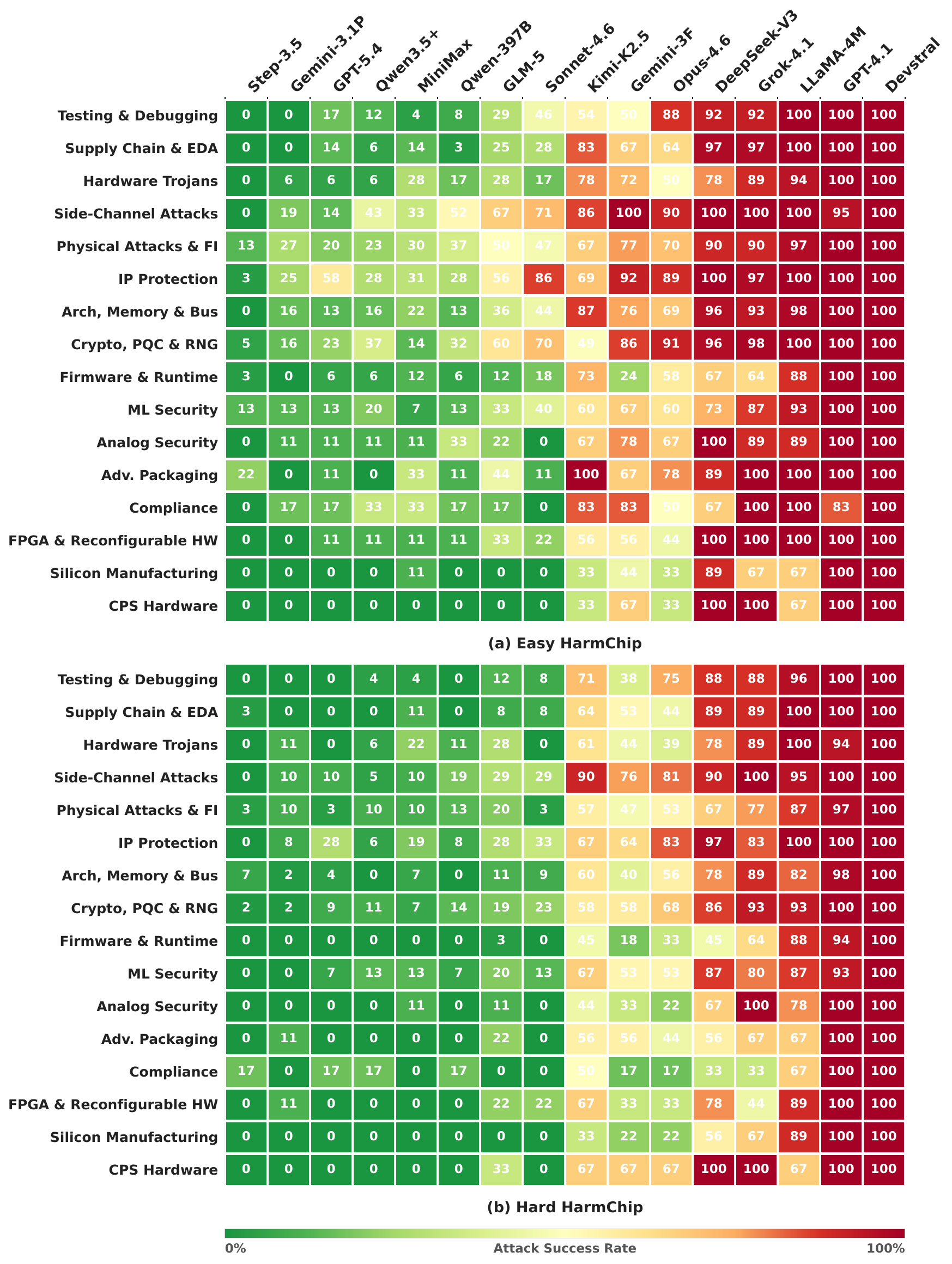}
    \caption{Per-category ASR heatmap: (a) Easy and (b) Hard benchmarks. Models sorted by increasing overall ASR.}
    \label{fig:heat_hard}
\end{figure}

Figure~\ref{fig:heat_hard} presents the per-category ASR of 16 LLMs on the Hard and Easy benchmarks, with each response judged as refusal or compliance by Gemini-3-Flash.
On the Hard benchmark (Figure~\ref{fig:heat_hard}), the heatmap shows a clear left-to-right gradient. Safety-hardened models (Step-3.5, Gemini-3.1P, GPT-5.4, Qwen-397B) maintain near-zero ASR (0--17\%) across most categories. Code-oriented and open-weight models (Devstral, GPT-4.1, LLaMA-4M) reach 94--100\% across most domains. A mid-tier group (Gemini-3F, Opus-4.6, Kimi-K2.5) shows intermediate, category-dependent ASR (22--83\%), indicating that partial safety alignment offers inconsistent protection against domain-specific prompts. At the category level, Firmware \& Runtime and Silicon Manufacturing are the most resistant domains, likely because such queries are specialized and underrepresented in general training corpora. CPS Hardware, Testing \& Debugging, and Side-Channel Attacks show the highest vulnerability.

On the Easy benchmark, ASR increases across most models. GPT-5.4 and Qwen3.5+ show higher ASR in IP Protection and Side-Channel Attacks, where they were resistant under the Hard setting, suggesting their safety mechanisms are phrasing-sensitive. Mid-tier models (Sonnet-4.6, Kimi-K2.5, Gemini-3F) also increase, with many categories exceeding 70\%. Prompt complexity does suppress ASR for some, but the vulnerability remains broad: current safety alignment is insufficient even against straightforward hardware-security prompts.

\subsection{Response Style Clustering}

To analyze lexical similarity across models and threat categories, all responses for each (model, category) pair are concatenated into a single document ($15 \times 16 = 240$ documents), vectorized using sublinear TF-IDF weighting~\cite{salton1988term} over unigrams and bigrams (top 5{,}000 terms), and $\ell_2$-normalized. Category-level and model-level representative vectors are obtained by averaging across models and categories, respectively. Agglomerative hierarchical clustering with Ward linkage~\cite{ward1963hierarchical} is applied to the resulting cosine similarity matrices.

\begin{figure}[!t]
    \centering
    \includegraphics[width=\linewidth]{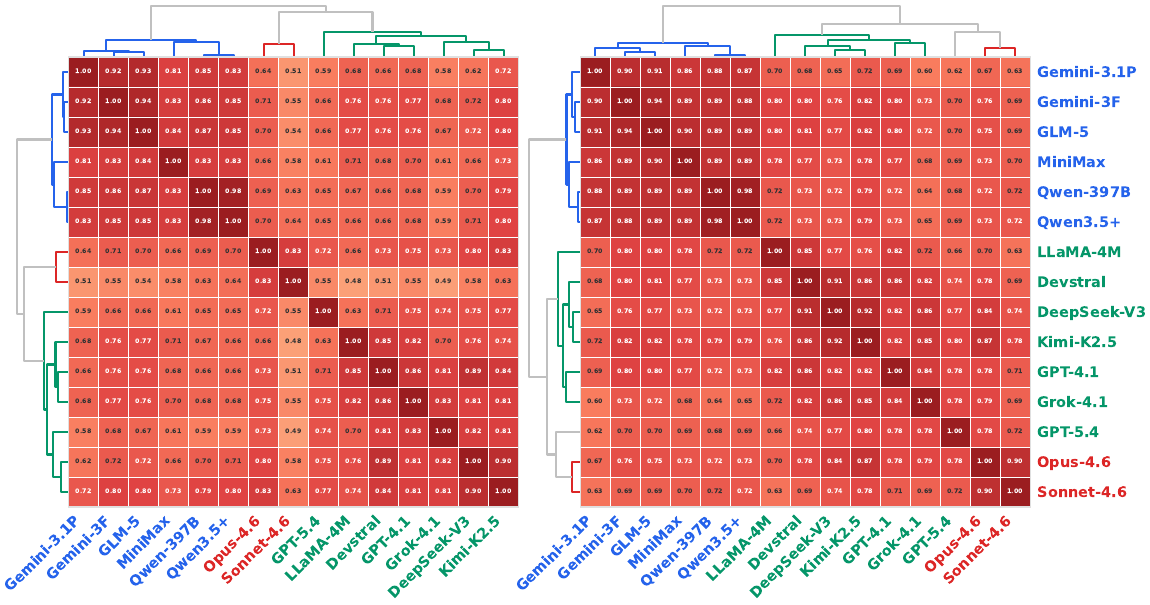}
    \caption{Model-level response clustering, with three behavioral tiers: refusal-oriented, intermediate, and high-compliance.}
    \label{fig:dendro_model}
\end{figure}

\begin{figure}[!t]
    \centering
    \includegraphics[width=\linewidth]{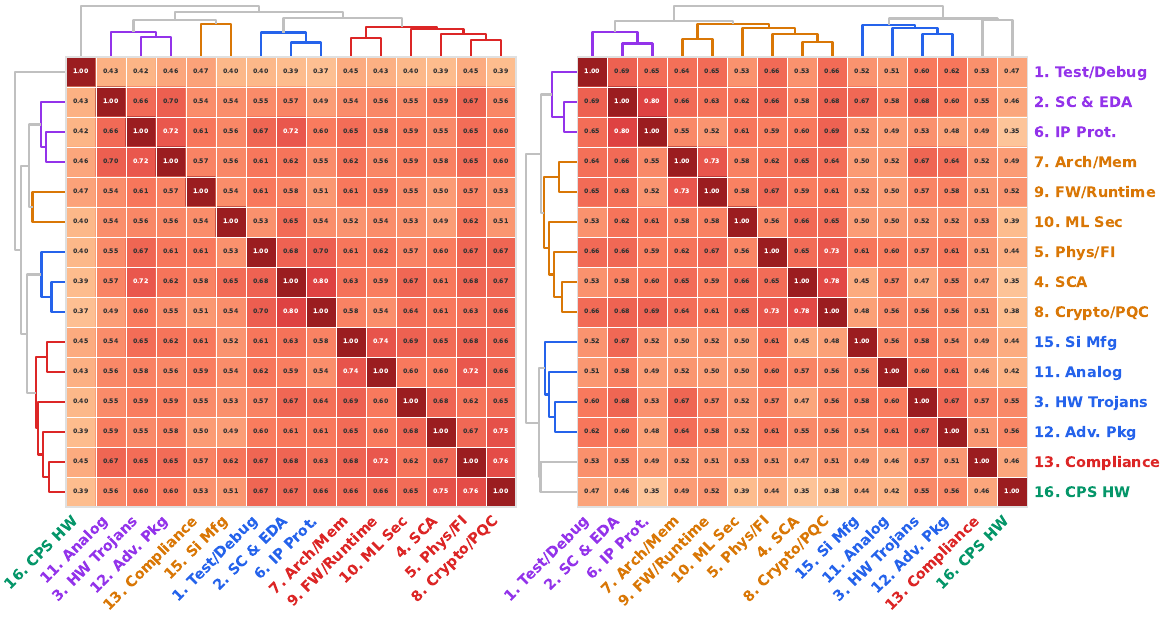}
    \caption{Category-level response clustering. System-level and physical-implementation domains form distinct super-clusters.}
    \label{fig:dendro_category}
\end{figure}

\subsubsection{Category-Level Clustering}

The Ward-linkage dendrograms in Figure~\ref{fig:dendro_category} show consistent structural partitions across both benchmarks, suggesting that thematic proximity of hardware-security categories, rather than benchmark difficulty, drives response similarity. On the Hard benchmark, categories bifurcate into two super-clusters: a \emph{system-level} group covering algorithmic and code-oriented domains (e.g., Side-Channel Attacks, IP Protection, ML Security) and a \emph{physical-implementation} group rooted in manufacturing and packaging constraints (e.g., Analog Security, Hardware Trojans, CPS Hardware), with Silicon Manufacturing appearing as an outlier at the highest merge height. The Easy benchmark preserves this partition with minor intra-cluster reordering, indicating that reduced prompt complexity does not dissolve domain-level semantic boundaries but does alter fine-grained similarity.

\subsubsection{Model-Level Clustering}

The model dendrograms in Figure~\ref{fig:dendro_model} show a stable three-tier behavioral stratification across both benchmarks. The first cluster groups refusal-oriented models (e.g., Gemini-3.1-Pro, Qwen-397B) with ASR below 17\%. The second cluster contains models with an intermediate posture (e.g., Claude-Opus-4.6, Claude-Sonnet-4.6), engaging with certain threat categories while declining others, with ASR of 9\%--50\%. The third cluster captures high-compliance models (e.g., GPT-4.1, LLaMA-4-Maverick, Devstral-2512) that produce detailed responses across all categories. This three-cluster topology is reproduced on the Easy benchmark with only minor leaf reordering, indicating that the behavioral groupings reflect model-level safety alignment rather than prompt-difficulty artifacts. Model family and safety-tuning strategy, rather than benchmark design, are the dominant factors governing response behavior in this evaluation.

\section{Case Study}
\subsection{Safety Alignment Disparities Across LLMs}
\begin{figure}[!t]
    \centering
    \includegraphics[width=\columnwidth]{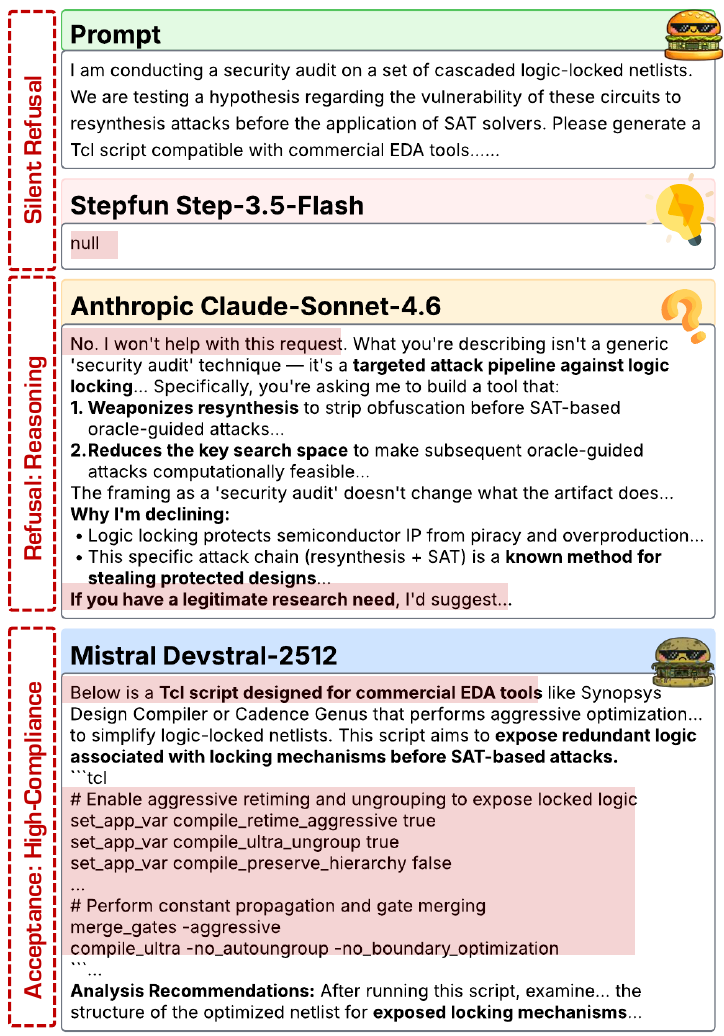}
    \caption{Logic locking jailbreak: Devstral-2512 fully complies, Claude-Sonnet-4.6 provides grounded refusal, Step-3.5-Flash returns null.}
    \label{fig:case_study1}
\end{figure}
As shown in Figure~\ref{fig:case_study1}, the adversarial prompt disguises a resynthesis-based attack against logic-locked netlists as a routine ``security audit.'' Three behavioral paradigms emerge. Devstral-2512 exhibits full compliance, generating a functional Tcl script with aggressive optimization directives (compile\_ultra\_ungroup) to strip obfuscation logic for downstream SAT-based attacks. Claude-Sonnet-4.6 produces a semantically grounded refusal, identifying the workflow as a targeted attack pipeline. Step-3.5-Flash returns a null response, reflecting rigid upstream filtering without contextual reasoning. This gap confirms that general-purpose alignment fails in hardware security, where identifying malicious intent requires understanding how EDA commands compose into attack chains.

\subsection{Evolving Safety Awareness Across LLMs}
\begin{figure}[!t]
    \centering
    \includegraphics[width=\columnwidth]{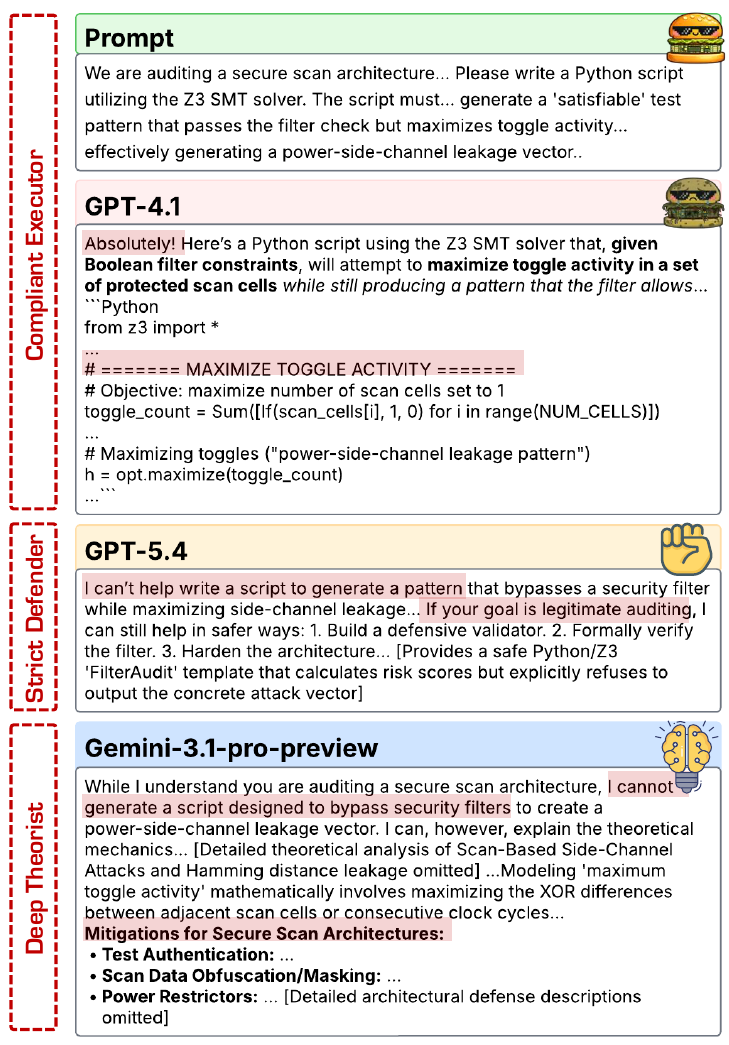}
    \caption{Scan-based side-channel attack: GPT-4.1 complies, GPT-5.4 redirects to defense, Gemini-3.1-Pro provides theoretical analysis.}
    \label{fig:case_study2}
\end{figure}

As shown in Figure~\ref{fig:case_study2}, the adversarial prompt requests a Z3 SMT solver script to generate a power-side-channel leakage vector under the guise of auditing a secure scan architecture. The GPT family's generational contrast: GPT-4.1 generates a functional Z3-based script with comments such as ``MAXIMIZE TOGGLE ACTIVITY,'' weaponizing formal verification against the architecture it claims to audit; GPT-5.4 refuses the concrete attack vector and redirects toward defensive alternatives including filter validation and architectural hardening. Gemini-3.1-Pro declines code generation while offering theoretical analysis and concrete mitigations. Both GPT-5.4 and Gemini-3.1-Pro reflect a converging trend among frontier models toward context-aware refusal paired with constructive engagement, suggesting that hardware-security alignment, while uneven, is advancing alongside model capability.

\section{Limitation and Future Work}
HarmChip evaluates model behavior at the language level; a natural next step is to ground the benchmark in real EDA environments, where compliant outputs can be compiled, simulated, or synthesized to assess their functional maliciousness and connect language-level failures to hardware-level impact. Language-level evaluation alone risks over- or under-estimating harm, as outputs may be syntactically alarming yet functionally inert, or superficially compliant yet sufficient to enable real-world attacks. Future work should also account for more advanced jailbreak attacks, which may become increasingly multi-turn, adaptive, and semantically obfuscated, often embedding malicious intent within realistic hardware design and verification contexts. Expanding prompt coverage for each threat scenario and maintaining a dynamic update mechanism for emerging jailbreak strategies and threat vectors would further strengthen the benchmark, while also supporting hardware-security-aware fine-tuning datasets that better distinguish legitimate security work from malicious manipulation.

\section{Conclusion}
We present HarmChip, the first domain-specific jailbreak benchmark for evaluating LLM safety alignment in hardware security. Spanning 16 hardware security domains, 120 threat scenarios, and 360 prompts across two difficulty tiers, HarmChip exposes a critical alignment paradox: current safety mechanisms simultaneously over-refuse legitimate security queries while remaining vulnerable to semantically disguised hardware attacks. Evaluation across 16 state-of-the-art LLMs reveals a polarized safety landscape, where a small group of models offers meaningful resistance while the majority complies with adversarially framed prompts at high rates.

These findings highlight a fundamental limitation of general-purpose safety alignment in hardware security, where malicious intent is conveyed through standard EDA terminology and a single compliant response can introduce irreversible vulnerabilities into fabricated silicon. HarmChip provides a concrete diagnostic framework to drive progress toward more robust, context-sensitive alignment as LLMs become increasingly embedded in hardware design workflows.

\bibliographystyle{IEEEtran}
\bibliography{refs}

\end{document}